\documentclass[a4paper,fleqn,usenatbib]{mnras}

\usepackage{mathptmx}
\usepackage{subfig}

\usepackage[T1]{fontenc}
\usepackage{ae,aecompl}

\usepackage{graphicx}	
\usepackage{amsmath}	
\usepackage{amssymb}	

\title[Morphological classification of radio galaxies]{Morphological classification of radio galaxies: Capsule Networks versus Convolutional Neural Networks}

\author[V. Lukic et al.]{
V. Lukic,$^{1}$\thanks{E-mail: vesna.lukic@hs.uni-hamburg.de}
M. Br\"uggen,$^{1}$\thanks{E-mail: mbrueggen@hs.uni-hamburg.de}
B. Mingo,$^{2}$
J. H. Croston,$^{2}$
G. Kasieczka,$^{3}$
P.N. Best$^{4}$
\\
$^{1}$ Hamburg Observatory, University of Hamburg, Gojenbergsweg 112, Hamburg 21029, Germany \\
$^{2}$ School of Physical Sciences, The Open University, Walton Hall, Milton Keynes, MK7 6AA, UK \\
$^{3}$ Institute of Experimental Physics, University of Hamburg, Luruper Chaussee 149, 22761 Hamburg \\
$^{4}$ Institute for Astronomy, University of Edinburgh, Royal Observatory, Blackford Hill, Edinburgh EH9 3HJ, UK
}
\date{Accepted XXX. Received YYY; in original form ZZZ}

\pubyear{2019}

\begin{document}
\label{firstpage}
\pagerange{\pageref{firstpage}--\pageref{lastpage}}
\maketitle

\begin{abstract}
Next-generation radio surveys will yield an unprecedented amount of data, warranting analysis by use of machine learning techniques. Convolutional neural networks are the deep learning technique that has proven to be the most successful in classifying image data. Capsule networks are a more recently developed technique that use capsules comprised of groups of neurons, that describe properties of an image including the relative spatial locations of features. The current work explores the performance of different capsule network architectures against simpler convolutional neural network architectures, in reproducing the classifications into the classes of unresolved, FRI and FRII morphologies. We utilise images from a LOFAR survey which is the deepest, wide-area radio survey to date, revealing more complex radio-source structures compared to previous surveys, presenting further challenges for machine learning algorithms. The 4- and 8-layer convolutional networks attain an average precision of 93.3\% and 94.3\% respectively, compared to 89.7\% obtained with the capsule network, when training on original and augmented images. Implementing transfer learning achieves a precision of 94.4\%, that is within the confidence interval of the 8-layer convolutional network. The convolutional networks always outperform any variation of the capsule network, as they prove to be more robust to the presence of noise in images. The use of pooling appears to allow more freedom for the intra-class variability of radio galaxy morphologies, as well as reducing the impact of noise.
\end{abstract}




\section{Introduction}
\label{sec:Intro}

Active Galactic Nuclei (AGN) are energetic, astrophysical sources powered by accretion onto super-massive black holes in galaxies \citep{Padovani_2017,Fabian_1999}. There are many classes of AGN, where one subset is radio-loud AGN, also known as radio galaxies. The two main ways of classifying radio galaxies is by the properties of optical emission lines \citep{Hine_Longair_1979} or by the radio morphology of the jets \citep{Bicknell_1995}. The classification of radio galaxy morphology is of research interest in wide-field radio surveys as it correlates with physical properties of the galaxy such as the total power, dust distribution, surrounding environment, and galaxy and cluster evolution \citep{Saripalli_2012}. Radio galaxies can present compact or extended radio morphologies \citep{Miraghaei_Best2017} and are often classified into either the FRI (core-bright) or FRII (edge-bright) galaxies \citep{Fanaroff_1974}. Rarer are hybrid galaxies, which fall in between FRI and FRII galaxies \citep{Gopal-Krishna_Witta2000}. There are physical differences between the two classes. The jets of FRIs are less powerful, and are disrupted quite close to the core of the radio galaxy, while the jets of FRII are more powerful and stay relativistic for much larger distances, terminating in a shock \citep{Contopoulos_etal_2015}. The transition from FRII to FRI radio galaxies is thought to occur as the jet becomes sub-relativistic \citep{Bicknell_1994}. As the environment plays a large role in the morphology of radio galaxies, it is not unusual for both lobes to have different appearances, especially the FRIs. The dynamics of the ambient gas and the motion of the host galaxy can create tails or distort the jets through ram pressure stripping \citep{Feretti_2003}. Compact radio sources may be either scaled-down (young) versions of the FRI or FRII sources, or may represent a physically distinct population \citep{Baldi_etal_2015}.

Radio surveys map ever-increasing numbers of radio sources. The visual classification of such sources becomes increasingly time-consuming and will be completely unfeasible with the rapidly increasing data volumes. Recent and upcoming surveys, such as the LOFAR Two-Metre Sky Survey \citep[LoTSS;][]{Shimwell_etal2017}, the Evolutionary Map of the Universe \citep[EMU;][]{Norris_2011} and surveys with the Square Kilometre Array \citep[SKA;][]{Prandoni_Seymour2015} will detect many millions of galaxies. Citizen science projects have been used for classifying astronomical sources, for example in Galaxy Zoo 2 \citep{Willett_etal2013} and Radio Galaxy Zoo \citep{Banfield_etal2015}. It is also possible to use automated techniques to classify images. Ultimately, these approaches can be used as a training set for machine learning algorithms, in particular deep learning algorithms, when the data is high-dimensional \citep{Wu_etal2018}.

The most prominent wide-area radio surveys, such as the Faint Images of the Radio Sky at Twenty centimetres \citep[FIRST;][]{Becker_1995} and the NRAO VLA Sky Survey \citep[NVSS;][]{Condon_etal_1998}, have mostly been conducted at GHz frequencies. In contrast, the LoTSS survey, which is the focus of the current work, has been carried out at 150 MHz with the Low Frequency Array (LOFAR). As such, LOFAR can detect synchrotron emission from older populations of relativistic electrons (which have steeper spectra) found in the extended regions  of sources. Furthermore, with its combination of long and short baselines, LoTSS offers both a high angular resolution ($\approx 6$\arcsec) for detailed mapping, and a high sensitivity to extended emission.

The cross-identification of radio sources with their optical or infrared hosts helps to associate radio components to sources and to determine properties, such as host galaxy redshift and mass. Previously, cross-identification has been done using visual input from citizen scientists input in Radio Galaxy Zoo \citep{Banfield_etal2015}, and automated methods in cross-identifying radio emission with infrared counterparts have been explored \citep{Alger_etal2018}. In the LoTSS survey \citep{Shimwell_etal2019} the radio sources have been cross-matched with their optical counterparts. For the majority of sources a maximum-likelihood ratio test was adequate because the sources are small and unresolved. For  sources that are too large or complex, a visual host identification has been applied \citep{Williams_etal2019}.

The first published work on the automated image classification of radio sources using deep learning algorithms was \citet{Aniyan_Thorat2017} where they use a limited number of original radio galaxy images and apply aggressive augmentation to classify sources into FRI, FRII and bent-tailed classes. In previous work, we have shown that it is possible to classify radio sources into four categories based on the number of components belonging to the radio source and produced a classification accuracy of 94.8 \% \citep{Lukic_2018} on the Radio Galaxy Zoo (RGZ) DR1 catalogue (Wong et al, in prep). \citet{Alhassan_etal2018} developed a convolutional neural network model to classify FIRST sources into four classes including compact, FRI, FRII and bent-tail sources, achieving overall accuracies >90\%. \citet{Wu_etal2018} use regional convolutional networks to localise, recognise and classify sources, the best model obtaining a final mean average precision of 83.4\%, using the number of peaks and number of components of a particular radio source. This approach, however, does not always lend itself easily to clear morphological classifications in the FRI or FRII cases because the relative orientations of components are not taken into account. 

The aim of the current work is to compare the performance of two setups of deep learning networks (capsule networks and convolutional networks) in the classification of radio sources. As a data set, we used the first data release of the LoTSS survey \citep{Shimwell_etal2019}. Capsule networks are a more recently developed deep learning technique, invented to help preserve the local feature information within an image, which can be degraded in traditional convolutional networks, owing to the pooling operation. In the context of radio galaxies, the orientation and pattern of the emission is important as it determines the morphological classification. The data from the LOFAR LoTSS survey reveals sources in unprecedented detail, therefore one source that had a particular morphology in an earlier survey may be revealed to have a different one when imaged with LOFAR. 

This paper is outlined as follows: Section~\ref{sec:Dataset} describes the LOFAR dataset, including catalogue information and image data as well as how the classifications are generated. Section~\ref{sec:methods} discusses the pre-processing and augmentation applied to the original images. Section~\ref{sec:DeepLearning} describes the theory behind the two deep learning approaches explored, namely convolutional neural networks and capsule networks. Section~\ref{sec:results} explores the performance of different capsule network models against standard convolutional neural network setups, including transfer learning on the LOFAR data, when training on different sets of images. The results are also discussed in Section~\ref{sec:results}. Section~\ref{sec:conclusions} summarises our overall findings.

\section{LOFAR HETDEX v1.0 dataset}
\label{sec:Dataset}

\subsection{Source cutouts}
\label{sec:cutouts}

The sources in our dataset originate from a 424 square degree region of the HETDEX Spring Field, mapped from the LOFAR Two-metre Sky Survey (LoTSS), and release as Data Release 1 \citep{Shimwell_etal2019}. The LoTSS survey detects a total of 325,694 sources where the signal is five times that of the noise and the density of sources is a factor of approximately 10 times higher than the most sensitive existing very wide-area radio-continuum surveys. We use v1.0 of the value-added catalogue for the HETDEX-area data release of LoTSS. The first step in creating the value-added catalogue involved using PyBDSF\footnote{http://www.astron.nl/citt/pybdsf/} to produce a radio source catalogue for the field, after which a decision tree was used to further categorise the sources, with details provided in \citep{Williams_etal2019}. After filtering the 325,694 sources to only include those classified as resolved leaves 24,096 sources \citep{Shimwell_etal2019}. The catalogue also contains 180 columns describing the properties, such as redshift, position etc, of the sources. In order to exclude star-forming galaxies and sources with less certain redshift values, we made use of the AGN subsample of the LoTSS catalogue, derived by \citet{Hardcastle_etal2019} leaving 6708 sources. We note that this is a substantial limitation of the machine learning approach when using radio galaxy image data only, as it is generally not always possible to filter out the star-forming galaxies without the use of additional data at other wavelengths. The source classifications were only available for those 6708 sources classified as AGN and with known redshifts, therefore the analysis is restricted to this set. However, the accurate knowledge of redshift is not strictly required for morphological classification.

\begin{figure}
    \includegraphics[width=\columnwidth]{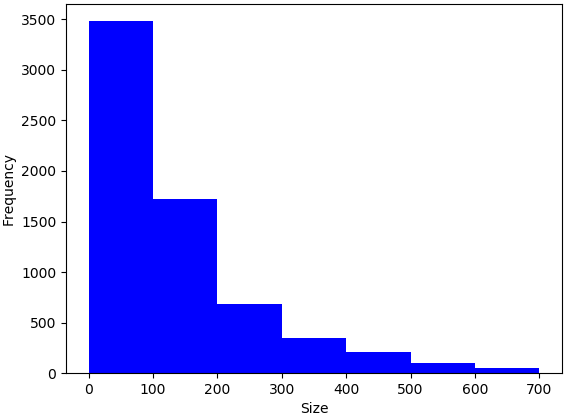}
    \caption{Histogram of sizes (in pixels per side) of the filtered cutout images. The total number of images is 6708.}
    \label{fig:shape_hist}
\end{figure}

Finally, we assume that there is one source per image. Square cutouts of each source are produced from the {\sc fits} images, where the cutout size is determined by the catalogued size of the radio source. These range from size (66,66) pixels up to (2342,2342) pixels. The size of the pixels is roughly 1.5x1.5\arcsec. Figure~\ref{fig:shape_hist} shows the histogram of the side length in pixels of the images for these 6708 samples.

\subsection{Classifications}
\label{sec:Classifications}
The LoTSS association and cross-identification effort \citep{Williams_etal2019} was a project in which expert astronomers were tasked with characterising the radio emission for sources larger than 15$''$. Indicated were the locations of the peaks and extents of the emission, and whether there was one or more sources present. 

\begin{figure}
    \includegraphics[width=\columnwidth]{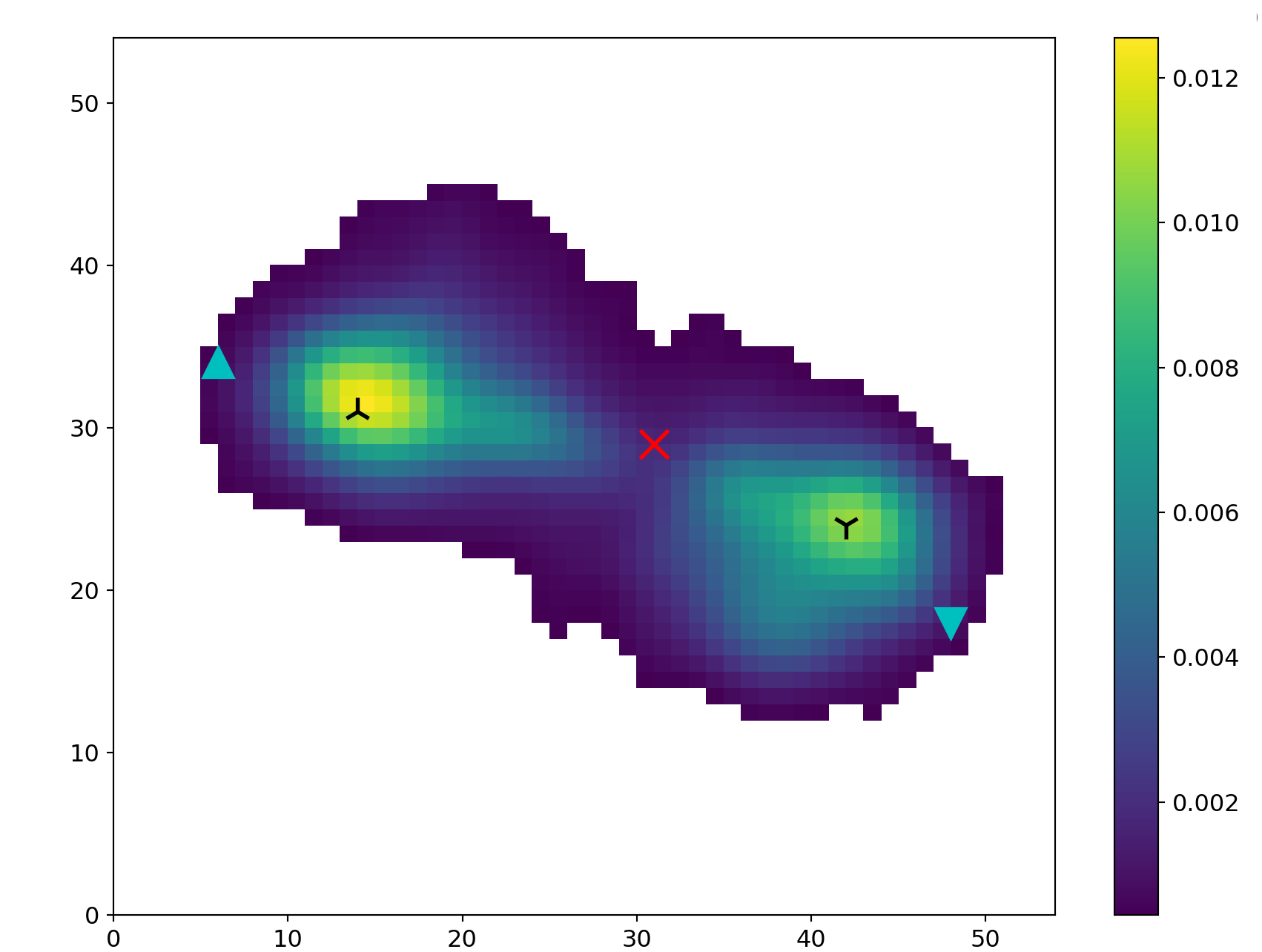}
    \caption{The masked array from which classifications are generated. The red cross indicates the position of the optical source, the black Y's indicate the peaks of the emission and the blue triangles indicate the maximum extents of emission. The optical position is calculated from the user's clicks on the LOFAR Two-metre Sky Survey images, or from the maximum likelihood method. The Y's and blue triangles are outputs from the automated classification code.}
    \label{fig:classifications_image}
\end{figure}

The 6708 source sample (see Section~\ref{sec:cutouts}) were classified into 6 classes using an automated technique (Mingo et al. in prep). The 6 classes are Unresolved-1, FRI, FRII, Hybrid-1, Hybrid-2 and Unresolved-2, all of which are described in further detail as follows. After the host galaxy location had been identified through the LoTSS identification effort \citep{Williams_etal2019}, the distances, d1 and d2, were determined as the distances in pixels from the host galaxy to the brightest peaks of emission on both sides of the source (shown with points marked with Y/inverted Y in Figure \ref{fig:classifications_image}). Similarly, Maxd1 and Maxd2 were determined as the maximum extents of the source in each direction (marked with triangles on the plots), out to the masked 4rms limit. A 120 degree aperture cone is used to find those along the direction of d1, d2. The comparison of d1/Maxd1 and d2/Maxd2 is then used to classify the sources. If, on both sides, the peak is less than half of the distance between the position of the host galaxy and the maximum extent of the emission (ie. d1/Maxd1 < 0.5 and d2/Maxd2 < 0.5) then the source is classified as an {\it FRI}, making up 15\% of the total sources. Likewise, if it is more than half of the distance (d1/Maxd1 > 0.5 and d2/Maxd2 > 0.5) then the source is classed as an {\it FRII}. The {\it FRIIs} make up 7\% of the total sources.

In addition to the FRI and FRII labels, four further labels were defined. {\it Hybrid-1} and {\it Hybrid-2} classes refer to sources which show FRI morphology on one side of the source and FRII in the other, with the `1' or `2' reflecting the classification of the brighter of the two sides. The {\it Hybrid} classes together make up 6\% of the sources. {\it Unresolved-1} sources correspond to those images that have less than 5 pixels of signal above 4rms, making up 22\% of the sources. This class is useful as it indicates which images are too noisy to be characterised into a particular class (note that it is different from the Unresolved sources previously discussed, which were based on the extent of the overall radio emission).  Finally, the {\it Unresolved-2} class contains a collection of mostly FRI and FRII sources that were unable to be classified accurately by the automated algorithm as they were too small, which makes up 50\% of the sources.  Figure~\ref{fig:classifications_image} shows an example image source, demonstrating how the classification labels were generated. 

In the current work, we have chosen the Unresolved-1 (henceforth called Unresolved), FRI and FRII classes to evaluate the performance of our deep learning algorithms, as these had the most confident classifications. There are 2901 original images in total, as shown in Table~\ref{tab:Number_sources_initial_augmented}.

The automated classification technique (Mingo et al. in prep) involved using masked 4rms arrays (where emission below 4rms is removed and potential unassociated emission is masked), rather than the raw {\sc fits} data. We define unassociated emission as radio emission which does not appear to belong to the radio source in question. A flood-filling algorithm\footnote{http://scikit-image.org/docs/dev/api/skimage.measure.html \\ \#skimage.measure.label} and masking techniques have additionally been applied in order to identify and use associated structures and consequently remove unassociated emission from the image (Mingo et al. in prep). On the other hand, the current work emphasises using the raw {\sc fits} images as the input to the deep learning algorithms, to see if they could be trained to cope with unassociated emission and unfiltered noise. After visual inspection we found there were approximately 1\% of images containing potentially unassociated emission, whereas the majority of the images contain varying levels of noise. 

In cases where the calibration did not perform as expected, the source will not be de-convolved accurately, causing flux leakage. This could result in the source being misclassified, leading to label errors. After inspecting several batches of images, we estimated the amount of labels containing errors to be less than 6\%, when considering both FRIs and FRIIs. Since larger sources are easier to classify, there is a decreased likelihood that they will be mislabeled, therefore the size of the source affects the presence of noisy labels. However, pre-filtering is applied to ensure the effect is not very large.

\begin{figure}
    \includegraphics[width=\columnwidth]{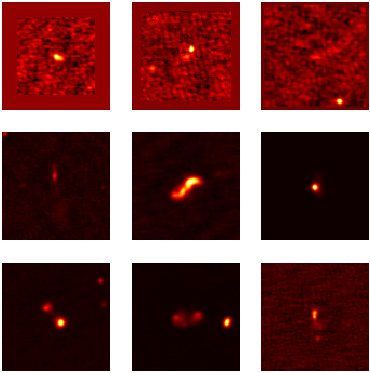}
    \caption{Showing morphology samples of the {\sc fits} cutouts when converted to png images using the \textquoteleft hot' colormap. The top row shows the \textquoteleft Unresolved class', middle row shows the FRI class, bottom row shows FRII. There are varying levels of noise and the occasional potentially unassociated emission present in the images.}
    \label{fig:fits_0_1_2}
\end{figure}

Figure~\ref{fig:fits_0_1_2} shows typical examples of source types across the three classes. It is evident that there are varying levels of noise present in the images, presenting the largest hindrance to the deep learning algorithms' ability to classify the sources accurately. One of the aims of the current work is to see how well the algorithms can classify the sources in the presence of such undesirable features, present in the original radio images ({\sc fits} files). We also compare the results obtained when using the masked 4rms clipped arrays (see Section \ref{sec:Sigma-clipped}), where emission below 4rms is removed and potential unassociated emission is masked. 

\section{Methods}
\label{sec:methods}

We use the radio galaxy image {\sc fits} cutouts from version $1.0$ of the LoTSS DR1 value-added catalogue \citep{Williams_etal2019}. The extended source identifications do not differ from the final version to a large extent. 

\subsection{Pre-processing}

Since the size of each cutout varies, they first need to be made the same size. The {\sc fits} images have been resized to (200,200) pixels, where the smaller images have been padded with zeros around the edges, and the larger images have been downsampled, using bicubic interpolation. The sizes of the arrays varies across all three classes. Following this, the images are centred on the position of the optical source, ensuring its position is at (100,100). We crop to the inner (100,100) pixel part of the image as the source is likely to be contained in this interval and to reduce the amount of data input into the network. The pixel values, representing brightness in mJy/beam were normalised by dividing by the maximum value in each image, therefore the values are contained within the [0,1] range. The images are taken at 150MHz. We apply the  \textquoteleft hot' colormap from the python matplotlib library, which converts the images from a single channel numpy array to a RGB png image. This is done by assigning a color (RGB vector) according to the value in the single channel array. For example, values close to 1 are bright yellow in the \textquoteleft hot' colormap scheme, therefore (r,g,b) $\approx$ (1,1,0.99). The conversions to the RGB vector are provided\footnote{y =(0,0.36): (r,g,b) $\approx$ (x=y/0.36,0,0) \\ y = (0.36,0.74): (r,g,b) $\approx$ (1,x=(y-0.37)/0.37,0) \\  y = (0.74,1): (r,g,b) $\approx$ (1,1,x=(y-0.75)/0.25)}. The conversion is done to make the arrays more amenable to deep learning analysis and has no bearing on the flux values. The number of sources in each class is given in Table~\ref{tab:Number_sources_initial_augmented}. 

Cropping the images to (100,100) pixels, instead of using the originally resized images of (200,200) pixels, reduces the impact of radio emission that is potentially unassociated with the main source in the centre. We have also experimented with using central sizes other than (100,100) pixels, however they resulted in worsened performance metrics. Smaller images tended to have some associated emission truncated, whereas larger images encapsulated more unassociated emission. The cropping still preserved the general noise characteristics surrounding the source.

The upsizing of images should not have any detrimental effects on image quality, however the downsizing may cause effects such as as slight distortion of the radio emission due to the interpolation. 

\begin{table}
	\centering
	\caption{The number of original and augmented sources, divided into training and testing sets. The percentage of samples in each class is also given for the test set. Since only original images should be used in the test set, the augmented images are used for training only.}
	\label{tab:Number_sources_initial_augmented}
	\begin{tabular}{lcccr}
		\hline
		Class & \# Orig.(Train) & \# Orig.(Test) & \# Aug. & \# Total \\
		\hline
		Unres. & 1156 & 301 (50.2\%) & 4371 & 5828 \\
		FRI & 765 & 219 (36.5\%) & 5904 & 6888 \\
		FRII & 380 & 80 (13.3\%) & 2760 & 3220 \\
		\hline
		\hline
		Total & 2301 & 600 & 13035 & 15936 \\
		\hline
	\end{tabular}
\end{table}

\subsection{Image augmentation}
\label{sec:augmentation}

Deep learning algorithms generally require large numbers of labeled images in order to make predictions more successfully and to reduce the effect of overfitting, in which the algorithm memorises the training samples and therefore the model fails to generalise on an independent dataset. More images can be generated artificially, by performing simple transformations to the original data \citep{Krizhevsky_etal2012}. As such, we apply translation, rotation and flipping to generate more images. In using translation, we initially use a random number that shifts the image between 0 and 20 pixels in any of the four directions, using the condition that if such a translation moves the brightest pixel out of the image, the translation is reduced to 10\% of the original value. This is to reduce the possibility that part of a radio component will be shifted out of the image. The images have been rotated randomly in multiples of 90 degrees only in order to avoid interpolation artefacts. We note that since there is a limited range of rotation applied, it is not enough to ensure complete rotational invariance in our models. Both horizontal and vertical flipping has been applied at random. The augmentation of the FRI and FRII sources has been done keeping their overall proportions similar in number to the original dataset as this resulted in improved performance. The number of original and augmented images used in the current work is given in Table \ref{tab:Number_sources_initial_augmented}. Image augmentation is applied on both the original LOFAR images, as well as the masked 4rms arrays.

\section{Deep Learning algorithms}
\label{sec:DeepLearning}

The most successful class of machine learning methods in the context of extracting information from high-dimensional data is deep learning, which has achieved unprecedented performance in a variety of domains such as image recognition, sentiment analysis and genomics \citep{LeCun_Bengio_Hinton2015}. Their ability to learn multiple representations of data lies in their stacked layer architecture.
The most commonly used implementation of deep learning has to date been convolutional neural networks. However, more recent advances were made in addressing the lack of rotational invariance in convolutional neural networks through the development of capsule networks.

\subsection{Convolutional Neural Networks}
\label{sec:dnn_theory}

Neural networks and deep learning algorithms are generally trained using the backpropagation algorithm, where a gradient descent optimisation algorithm is used to minimise the error between the predictions
of the network and the input labels by calculating the gradients and adjusting the weights accordingly \citep{Rumelhart_etal1986}. A deep fully connected neural network becomes time-consuming and computationally intensive to train. Convolutional neural networks employ smaller sized filters that scan across the image and extract features, which greatly reduces the dimensionality compared to using adjacent layers of fully connected neurons and enforces parameter sharing and therefore translational invariance \citep{Karpathy2016}. Spatial pooling layers are typically inserted between at least one convolutional layer which further reduces the dimensionality of features propagated through the network. In max pooling, the maximum value of a certain region of the image is output into the next layer. However, since the pooling operation summarises the information in a local part of the image, the global feature information within the image tends to degrade.

\subsection{Capsule networks}
\label{sec:capsule_theory}

Capsule networks \citep{Sarbour_etal2017} have been developed to preserve the relative locations of features within images and thus model the hierarchical relationships better. Whereas traditional neural networks output a single activation value, capsule networks are higher dimensional and output a vector representing a group of parameters such as orientation, skew, thickness etc., depending on the input. The overall
length of these vectors give the probability that the entity exists. Capsule networks have achieved state of the art performance on the MNIST dataset \citep{LeCun_1998} without data augmentation \citep{Xi_etal2017}.

In the context of radio galaxy classification, capsule networks should be able to preserve the emission pattern features over a large spatial extent, given an adequate training set size. 

Below we summarize the theory behind capsule networks but see \citet{Sarbour_etal2017} for a detailed description. For all capsules above the first layer of capsules, the input to a capsule $s_j$ is a weighted sum over all prediction vectors from the capsules in the layer below, given by multiplying the coupling coefficients $c_{ij}$ by the output $u_i$ of a capsule in the layer below by a weight matrix $W_{ij}$, as shown in Equation~\ref{eq:1}

\begin{equation}
\label{eq:1}
s_j=\sum_i c_{ij}W_{ij}u_i
\end{equation}

The coupling coefficients $c_{ij}$ are determined by a routing softmax function given by Equation~\ref{eq:3}  

\begin{equation}
\label{eq:3}
c_{ij}=\frac{e^{b_{ij}}}{\sum_k e^{b_{ik}}}
\end{equation}
 
The coupling coefficient $c_{ij}$ is the level of agreement between the predicted output of capsules in a layer, to their parent capsules in the layer above. $b_{ij}$ gives the log prior probabilities that capsule $i$ should be coupled to capsule $j$.
 
The vector length is calculated as shown in Equation~(\ref{eq:2})

\begin{equation}
\label{eq:2}
v_j=\frac{\lvert{\lvert{{s_j}}}\rvert{}\rvert{}^2}{1+\lvert{\lvert{{s_j}}}\rvert{}\rvert{}^2}\frac{s_j}{\lvert{\lvert{{s_j}}}\rvert{}\rvert{}} ,
\end{equation}
where $v_j$ is the vector output of capsule $j$ and $s_j$ is its total input. This output gives the probability that a specific property exists in the
input to the capsule, that is represented by the capsule. The vector output $v_j$ is an activation function, that is also referred to as a 
squashing function as it shrinks short vectors to near zero if a property is not present in the capsule, and long vectors to lengths close to 1 if 
the property exists. 

The agreement $a_{ij}$ for updating log probabilities and coupling coefficients is given by Equation~(\ref{eq:4})

\begin{equation}
\label{eq:4}
a_{ij}=v_j.W_{ij}u_i
\end{equation}
A margin loss function is used in order to determine whether a radio galaxy of a particular class is present, which has the form given by Equation~(\ref{eq:5}):

\begin{equation}
\label{eq:5}
L_k=T_k\;max(0,m^+ - \lvert{\lvert{{v_k}}}\rvert{}\rvert{})^2+\lambda(1-T_k)\;max(0,\lvert{\lvert{{v_k}}}\rvert{}\rvert{}-m^-)^2 ,
\end{equation}
where $T_k=1$ if a radio galaxy of class $k$ is present and $m^+=0.9$ and $m^-=0.1$, to ensure that the vector length remains within reasonable bounds. The $\lambda$ down-weighting function is introduced for numerical stability and suggested to be set at 0.5.

The mean squared error difference between the reconstructed image from the decoder (the part of the Capsule network after LabelCaps) and the input image acts as a regulariser for the capsule network, such that
near-perfect reconstructions will produce a near-zero error and poor reconstructions will produce a large error. The reconstruction loss is
scaled down by 0.0005 so it does not dominate the margin loss during training, and the coefficient for the default model is designed
for the MNIST digits which have an image size of 28x28, thus the coefficient is worked out to be $0.0005 \times 28 \times 28=0.392$.

\begin{figure*}
    \includegraphics[width=17.5cm]{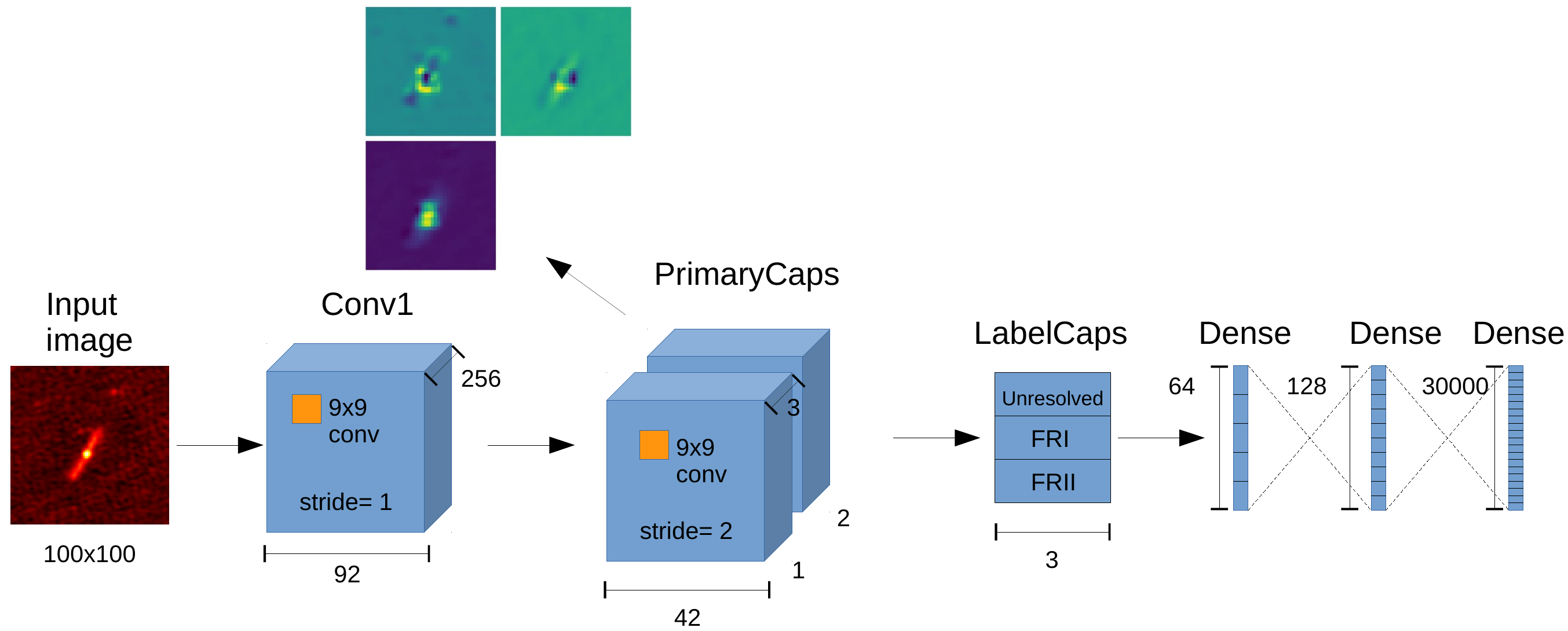}
    \caption{The default architecture for CapsNet, using three classes. The input to the network is a 100x100x3 image. The encoder is the part of the network that encapsulates the convolutional layer up to and including the LabelCaps layer. The decoder refers to the final three dense layers. An example of features detected by the PrimaryCaps layer prior to reshaping and squashing is shown, for the given input image. There is a small amount of extended emission to the top right of the image that appears to be unassociated with the main source in the centre, which the capsule network preserves, suggesting that it is not robust to potential unassociated sources. Additionally, the feature maps appear to show extra distortion in the core of the source.}
    \label{fig:capsnet_architecture}
\end{figure*}

\begin{table}
	\centering
	\caption{Showing architecture for the default capsule network model}
	\label{tab:CapsNet_architecture}
	\begin{tabular}{lcr}
		\hline
		Layer & Output shape & \# Params \\
		\hline
		Input\_1 & (None, 100, 100, 3) & 0 \\
		conv2d & (None, 92, 92, 256)& 62464 \\
		PrimaryCap\_conv2d & (None, 42, 42, 6)  & 124422 \\
		PrimaryCap\_reshape & (None, 3528, 3)  & - \\
		PrimaryCap\_squash & (None, 3528, 3) & - \\
		LabelCaps & (None, 3, 3)  & 95256 \\
		Input\_2 & (None, 3)  & - \\
		mask &  (None, 9) & - \\
		capsnet & (None, 3)  & - \\
		decoder &  (None, 100, 100, 3) & 3878960 \\
		\hline
		Total & & 4,161,102 \\
		\hline
	\end{tabular}
\end{table}

\subsection{Deep learning parameters}

There are several deep learning implementations currently available for use. The present work uses 
Keras\footnote{https://keras.io/preprocessing/image/} with the TensorFlow\footnote{https://www.tensorflow.org} backend and Python version 2.7.14. 

We use the Adam optimiser \citep{Kingma_Ba2014} with the default learning rate of 0.001. In order to keep more parameters the same between 
the models, both the convolutional and capsule network models are trained using a batch size of 100, for 50 epochs. 

The deep-learning task is a multi-classification problem, where the models output a 3-dimensional vector representing the probability that the object belongs to each class. The predicted class is chosen as the one with the largest probability value. As the probabilities are independent, there is no constraint that they need to add to unity. 

The models are trained using CPUs from 27 available Intel XEON CPU nodes with six available cores per node on a computing cluster at the University of Hamburg.

\subsubsection{ConvNet-4 parameters}

We use an architecture of two pairs of stacked convolutional layers with pooling layers in between, as shown in Figure~\ref{fig:conv_feature_Avg. F1s}, with parameters given in Table~\ref{tab:Stacked_architecture1}. This model is referred to as ConvNet-4. Using two adjacent convolutional layers with smaller filter sizes obtained improved results compared to using a single larger convolutional layer, and also reduced the number of parameters \citep{Simonyan_Zisserman2015}. We use the categorical cross-entropy cost function\footnote{https://keras.io/losses/\#categorical\_crossentropy} and 16 filters of size 5x5 across all layers, as well as the default learning rate decay of 0.
In order to reduce the effect of overfitting, dropout layers are used. A dropout value of 0.25 is used after each pair of convolutional layers, and a value of 0.5 in between the dense layers. A penalty term is added to the cost function using L2 regularisation \citep{Ng_2004} in the first dense layer. All the convolutional layers use the ReLU activation function \citep{Nair_Hinton2010}, and the softmax activation function at the final layer where classifications are made. There are 5,022,467 trainable parameters in total.

\subsubsection{ConvNet-8 parameters}
In order to investigate the performance for more complex convolutional networks, we can add additional layers. The ConvNet-8 model uses an architecture of four pairs of stacked convolutional layers with pooling layers in between. There are also an increasing number of feature maps with each subsequent double stacking of convolutional layers, as shown in Table~\ref{tab:Stacked_architecture2}. The architecture also uses smaller feature maps of size 3x3. There are 7,446,259 trainable parameters in total.

\subsubsection{CapsNet parameters}

Finally, we explore several variations of capsule network models. We downloaded the original CapsuleNet\footnote{https://github.com/XifengGuo/CapsNet-Keras/blob/master/capsulenet.py} code implemented in Keras that was built for the MNIST dataset \citep{Sarbour_etal2017}, and modified the code to use our datasets, vary the models from the original architecture and to calculate the metrics. The original architecture contains approximately 58M parameters, which is more than 14x the number of parameters as for the ConvNet-4 model. We therefore simplified the architecture to one having just over 4M parameters, and refer to this as the default model. The original CapsuleNet model is simplified in order to have the same order of magnitude as the parameters in the ConvNets and to help prevent overfitting.

The default architecture of CapsNet and decoder is illustrated in Figure~\ref{fig:capsnet_architecture} and the number of parameters is given in Table \ref{tab:CapsNet_architecture}. In essence it is comprised of an encoder and decoder. The encoder consists of a convolutional layer, which extracts features in the image, which are then input into the first capsule layer (PrimaryCaps), whose function is to take the 256x9x9 output of the convolutional layer and produce combinations of the detected features. The output of the PrimaryCaps layer is then sent to the LabelCaps layer, which produces one 3D capsule for each of the three radio galaxy classes. Routing is used between the PrimaryCaps layer and the LabelCaps layer such that the level of agreement of feature existence can be quantified and contribute to the vector length of the capsule. The decoder refers to the part of the network after the LabelCaps layer (the three dense layers at the end). There are 4,161,102 free parameters in the default CapsNet model.
 
We use 256 filters in the first convolutional layer, a filter size of 9 in both the first Convolutional layer and PrimaryCaps layer, 3 capsules in the PrimaryCaps and LabelCaps layers, 2 channels in the PrimaryCaps and the decoder contains (64,128) nodes. We use the default setup of three routings and a learning rate of 0.001 with a decay of 0.9. The first convolutional layer uses the ReLU activation function. CapsNet has image augmentation built into the training of the model, which we disable in order to use our augmentation technique, that allows more control over which classes get augmented and the type of transformations that are used. For the default CapsNet model, there are 4,161,102 parameters, which is a very similar number of parameters that was used for ConvNet-4. 

In addition to the default CapsNet model, we experiment with two other CapsNet models. In the first of these models (Inc. filtersize), we set the filter size to 24 and 18 in the first Convolutional layer and PrimaryCaps layer respectively and slide the filters across using a stride of 4 in the convolutional layer. The inc. filtersize model has 4,819,470 parameters. In the second model (Inc. decoder), we increase the complexity of the decoder to (128,256) nodes in the dense layers and the loss function of the decoder weight is increased from 0.392 to 5 respectively. The weight is calculated by taking the scaled-down reconstruction loss and multiplying it by the size of the images $0.0005 \times 100 \times 100 = 5$. There are 8,026,446 parameters in the inc. decoder model.

We chose to increase the filters from a size of 9 pixels in the inc. filtersize model because the original filter sizes that were designed for the MNIST image sizes of (28,28) pixels are likely too small compared to what would be needed for our (100,100) pixel images. We also experimented with increasing the number of nodes and weight loss of the decoder in the inc. decoder model to better account for the noise and potential unassociated emission in the dataset, as well as more variability in and between classes.

\section{Results}
\label{sec:results}

Due to the inherent stochasticity of training deep learning models, each run can produce slightly different results. We therefore train each model five times. The training data is also shuffled for each run to ensure there is no correlation between subsequent samples. There are several classification metrics that can help evaluate the performance of a classifier. In imbalanced class problems, the classification accuracy alone has several weaknesses in distinguishing between the performance of models \citep{Hossin_Sulaiman_2015}. The precision, recall and F1 scores are more informative measures of performance compared to using the classification accuracy. Precision refers to the fraction of true positives returned among all returned positive instances, recall is the fraction of true positives that are identified correctly, which also gives an indication of the sensitivity of the classifier. The F1 score is the harmonic mean of precision and recall, and can be interpreted as the average of the precision and recall values. The accuracy is the total proportion of correct predictions. Precision, recall, F1 score and accuracy are defined in Eqs.~(\ref{eq:x})-(\ref{eq:x2}).  

\begin{align}
\label{eq:x}
\mathrm{Precision} = \frac{\mathrm{TP}}{\mathrm{TP+FP}} \\
\mathrm{Recall} = \frac{\mathrm{TP}}{\mathrm{TP+FN}} \\
\mathrm{F1\_score} = \frac{2 \times \mathrm{Precision}\times \mathrm{Recall}}{\mathrm{Precision+ Recall}} \\ 
\label{eq:x2}
\mathrm{Accuracy} = \frac{TP+TN}{TP+FP+TN+FN} ,
\end{align}
where TP refers to the true positives, FP refers to the false positives and FN refers to false negatives. A true positive is when the prediction matches the label. A false positive is when the positive class is incorrectly predicted. A false negative is when the positive class is predicted to be in another class.

We also calculate the $95\%$ confidence interval using the mean and standard deviation of the metrics to account for the variability in performance across the runs. We declare a model to be statistically significantly better than another model if the mean of its metrics is higher than the 95\% confidence interval of the other models metrics. In order to ensure a fair comparison, the same training and testing sets were used for the ConvNet and CapsNet architectures.

The same set of data is used for both validation and testing when running the models, with the exception of the application of early stopping (results shown in Section \ref{sec:Early_stopping}). When early stopping is used, the validation data is used to determine when to stop the training. Otherwise, the use of the same dataset for validation and testing is of no consequence, as the weights that are modified using the training set are applied to the validation/test set to calculate the loss. No adjustment is made to the weights using the validation set. At the conclusion of training, the final weights are applied to the validation/test set and the metrics are calculated.

Section \ref{sec:LOFAR_orig} of the results shows the classification metrics across the two deep learning techniques when using the original data only, with 2301 (79\%) samples for training, and 600 (21\%) samples for both validation and testing. The fraction of samples in each class is given in Table \ref{tab:Number_sources_initial_augmented} for the test set. Section \ref{sec:orig+aug} makes use of augmented images in addition to the original images and Section \ref{sec:Sigma-clipped} explores the effects when the 4rms sigma-clipped data is used.

\subsection{LOFAR original images}
\label{sec:LOFAR_orig}
\subsubsection{ConvNet-4 and ConvNet-8 models}

We use the ConvNet-4 and ConvNet-8 models on the original 2901 images from LOFAR, which have been classified into Unresolved, FRI and FRII sources. 
The results are shown in Table~\ref{tab:LOFAR_conv_1} and Table~\ref{tab:LOFAR_conv_2}. Each epoch consisting of 2301 training samples takes approximately 32 and 66 seconds to train for ConvNet-4 and ConvNet-8 respectively.

\begin{table}
	\centering
	\caption{ConvNet-4 architecture. A filter size of 5 is used in the convolutional layers.}
	\label{tab:Stacked_architecture1}
	\begin{tabular}{lcr}
		\hline
		Layer & Output shape & \# Params \\
		\hline
		Input & (None, 100, 100, 3) & 0 \\
		conv2d & (None, 100, 100, 16) & 1216 \\
		conv2d & (None, 100, 100, 16) & 6416 \\
		maxpool2d & (None, 50, 50, 16)  & - \\
		dropout & (None, 50, 50, 16)  & - \\
		conv2d & (None, 50, 50, 16) & 6416 \\
		conv2d & (None, 50, 50, 16) & 6416 \\
		maxpool2d & (None, 25, 25, 16)  & - \\
		dropout & (None, 25, 25, 16) & - \\
		flatten & (None, 10000) & - \\
		dense & (None, 500) & 5000500 \\
		dropout & (None, 500) & - \\
		dense & (None, 3) & 1503\\
		\hline
		Total & & 5,022,467 \\
		\hline
	\end{tabular}
\end{table}

\begin{table}
	\centering
	\caption{ConvNet-8 architecture. A filter size of 3 is used in the convolutional layers.}
	\label{tab:Stacked_architecture2}
	\begin{tabular}{lcr}
		\hline
		Layer & Output shape & \# Params \\
		\hline
		Input & (None, 100, 100, 3) & 0 \\
		conv2d & (None, 100, 100, 32) & 896 \\
		conv2d & (None, 100, 100, 32)& 9248 \\
		maxpool2d & (None, 50, 50, 32)  & - \\
		dropout & (None, 50, 50, 32)  & - \\
		conv2d & (None, 50, 50, 64) & 18496 \\
		conv2d & (None, 50, 50, 64) & 36928 \\
		maxpool2d & (None, 25, 25, 64)  & - \\
		dropout & (None, 25, 25, 64) & - \\
		conv2d & (None, 25, 25, 128)  & 73856 \\
		conv2d & (None, 25, 25, 128) & 147584 \\
		maxpool2d & (None, 13, 13, 128)  & - \\
		dropout & (None, 13, 13, 128)  & - \\
		conv2d & (None, 13, 13, 256) & 295168 \\
		conv2d & (None, 13, 13, 256) & 590080 \\
		maxpool2d & (None, 7, 7, 256) & - \\
		dropout & (None, 7, 7, 256) & - \\
		flatten & (None, 12544) & - \\
		dense & (None, 500) & 6272500 \\
		dropout & (None, 500) & - \\
		dense & (None, 3) & 1503\\
		\hline
		Total & & 7,446,259 \\
		\hline
	\end{tabular}
\end{table}

\begin{table}
	\centering
	\caption{The average metrics (in percentages) across each of the classes in (1) the original LOFAR dataset, (2) the original and augmented dataset, (3) the original 4rms clipped dataset, and (4) the original and augmented 4rms clipped dataset for the ConvNet-4 model. Five runs were done in total, using 600 samples in the test set.}
	\label{tab:LOFAR_conv_1}
	\begin{tabular}{lcccr}
		\hline
		Class & Precision & Recall & F1 score & Accuracy \\
		\hline
		\textbf{(1)} & & & \\
		Unres. & 95.7 $\pm$ 0.9  & 96.7 $\pm$ 1.4  & 96.2 $\pm$ 0.9 & 95.9 $\pm$ 0.9 \\ 
		FRI & 86.2 $\pm$ 2.4 & 86.8 $\pm$ 1.1 & 86.5 $\pm$ 1.0 & 89.9 $\pm$ 0.9 \\
		FRII & 68.0 $\pm$ 1.1  & 63.5 $\pm$ 2.1  & 65.6 $\pm$ 1.0 & 90.9 $\pm$ 0.2  \\
		\hline
		\hline
		Avg. & 88.5 $\pm$ 0.8  & 88.7 $\pm$ 0.8 & 88.6 $\pm$ 0.9 & 93.1 $\pm$ 0.8 \\
		\hline
		\textbf{(2)} & & & \\
		Unres. & 98.1 $\pm$ 0.4  & 98.2 $\pm$ 0.5  & 98.1 $\pm$ 0.4 & 98.0 $\pm$ 0.4  \\ 
		FRI & 92.3 $\pm$ 0.9 & 93.3 $\pm$ 1.3  & 92.3 $\pm$ 0.2 & 94.2 $\pm$ 0.1 \\
		FRII & 80.9 $\pm$ 2.0 & 75.2 $\pm$ 4.9  & 77.8 $\pm$ 1.9 & 94.2 $\pm$ 0.2\\
		\hline
		\hline
		Avg. & 93.3 $\pm$ 0.2  & 93.4 $\pm$ 0.2 & 93.3 $\pm$ 0.2 & 96.2 $\pm$ 0.2 \\
		\hline
		\textbf{(3)} & & & \\
		Unres. & 97.9 $\pm$ 0.3 & 98.1 $\pm$ 0.5  & 98.0 $\pm$ 0.2 & 97.9 $\pm$ 0.2  \\ 
		FRI & 90.4 $\pm$ 0.7 & 90.0 $\pm$ 0.6  & 90.2 $\pm$ 0.4 & 92.8 $\pm$ 0.3 \\
		FRII & 72.1 $\pm$ 0.6 & 72.2 $\pm$ 1.6  & 72.1 $\pm$ 0.8 & 92.5 $\pm$ 0.2 \\
		\hline
		\hline
		Avg. & 91.8 $\pm$ 0.2  & 91.9 $\pm$ 0.3 & 91.8 $\pm$ 0.3 & 95.5 $\pm$ 0.2 \\
		\hline
		\textbf{(4)} & & & \\
		Unres. & 98.7 $\pm$ 0.6 & 99.7 $\pm$ 0.2  & 99.2 $\pm$ 0.2 & 99.2 $\pm$ 0.2  \\ 
		FRI & 91.5 $\pm$ 0.9 & 94.9 $\pm$ 0.6 & 93.1 $\pm$ 0.4 & 94.9 $\pm$ 0.3 \\
		FRII & 88.1 $\pm$ 1.3 & 75.5 $\pm$ 2.3  & 81.3 $\pm$ 1.4 & 95.3 $\pm$ 0.3 \\
		\hline
		\hline
		Avg. & 94.9 $\pm$ 0.2  & 94.7 $\pm$ 0.3 & 94.7 $\pm$ 0.2 & 97.3 $\pm$ 0.1 \\
		\hline
	\end{tabular}
\end{table}

\begin{table}
	\centering
	\caption{The average metrics (in percentages) across each of the classes in (1) the original LOFAR dataset, (2) the original and augmented dataset, and (3) the original 4rms clipped dataset, and (4) the original and augmented 4rms clipped dataset, for the ConvNet-8 model. Five runs were done in total, using 600 samples in the test set.}
	\label{tab:LOFAR_conv_2}
	\begin{tabular}{lcccr}
		\hline
		Class & Precision & Recall & F1 score & Accuracy \\
		\hline
		\textbf{(1)} & & & \\
		Unres. & 96.6 $\pm$ 1.1  & 98.8 $\pm$ 0.3  & 97.7 $\pm$ 0.5 & 97.5 $\pm$ 0.6  \\ 
		FRI & 88.7 $\pm$ 1.0 & 90.4 $\pm$ 0.7 & 89.6 $\pm$ 0.5 & 92.2 $\pm$ 0.4 \\
		FRII & 75.2 $\pm$ 4.1 & 64.5 $\pm$ 2.8  & 69.3 $\pm$ 1.1 & 92.3 $\pm$ 0.4 \\
		\hline
		\hline
		Avg. & 90.9 $\pm$ 0.4  & 91.2 $\pm$ 0.4 & 90.9 $\pm$ 0.5 & 94.9 $\pm$ 0.4 \\
		\hline
		\textbf{(2)} & & & \\
		Unres. & 98.2 $\pm$ 0.7  & 98.4 $\pm$ 0.2  & 98.3 $\pm$ 0.3 & 98.2 $\pm$ 0.3  \\ 
		FRI & 92.5 $\pm$ 0.6 & 94.0 $\pm$ 0.5  & 93.2 $\pm$ 0.4 & 95.0 $\pm$ 0.3 \\
		FRII & 84.5 $\pm$ 1.9 & 80.0 $\pm$ 1.0  & 82.2 $\pm$ 1.1 & 95.3 $\pm$ 0.3 \\
		\hline
		\hline
		Avg. & 94.3 $\pm$ 0.2  & 94.3 $\pm$ 0.2 & 94.3 $\pm$ 0.2 & 96.7 $\pm$ 0.1 \\
		\hline
		\textbf{(3)} & & & \\
		Unres. & 99.6 $\pm$ 0.3 & 98.8 $\pm$ 1.0  & 99.2 $\pm$ 0.5 & 99.1 $\pm$ 0.5 \\ 
		FRI & 92.7 $\pm$ 1.0 & 93.4 $\pm$ 3.3 & 93.0 $\pm$ 2.1 & 95.2 $\pm$ 1.7 \\
		FRII & 83.4 $\pm$ 9.3 & 83.4 $\pm$ 2.8  & 83.1 $\pm$ 5.8 & 95.2 $\pm$ 1.8 \\
		\hline
		\hline
		Avg. & 95.0 $\pm$ 1.6  & 94.9 $\pm$ 1.8 & 94.9 $\pm$ 1.7 & 97.3 $\pm$ 1.0 \\
		\hline
		\textbf{(4)} & & & \\
		Unres. & 99.6 $\pm$ 0.1 & 99.1 $\pm$ 0.4  & 99.3 $\pm$ 0.2 & 99.3 $\pm$ 0.2 \\ 
		FRI & 94.4 $\pm$ 0.4 & 95.2 $\pm$ 0.7 & 94.8 $\pm$ 0.4 & 96.2 $\pm$ 0.3 \\
		FRII & 86.0 $\pm$ 1.0 & 85.8 $\pm$ 1.5  & 85.9 $\pm$ 0.6 & 96.2 $\pm$ 0.1 \\
		\hline
		\hline
		Avg. & 96.0 $\pm$ 0.2  & 95.9 $\pm$ 0.2 & 95.9 $\pm$ 0.2 & 97.9 $\pm$ 0.2 \\
		\hline
	\end{tabular}
\end{table}

The models perform the best in recovering the images in the Unresolved class, which could be due to the images being generally noisier and the sources smaller, compared to the other images. The recovery of FRIIs is poorer however compared to the FRIs. This may be because there are fewer examples of images in this class (460 FRIIs compared to 984 FRIs). Although it can be argued that the morphological diversity is greater for the FRI class as they can be straight, bent, or one-sided with a peak at one end, FRIIs contain lobes that may or may not be connected, therefore the source can contain either one or two components. We have experimented with using different weights for the classes, giving proportionally greater weights for the FRIIs such that wrong predictions are penalised more, however the performance remained the same as before, across all classes. The recall (accuracy) tends to be higher compared to precision for the FRIs, whereas it is lower compared to precision for the FRIIs. This is likely due to it being easier to recover sources containing emission that is more concentrated in one place (in the case of the FRIs), compared to emission that is further apart. 

\begin{figure}
    \includegraphics[width=80mm,scale=0.9]{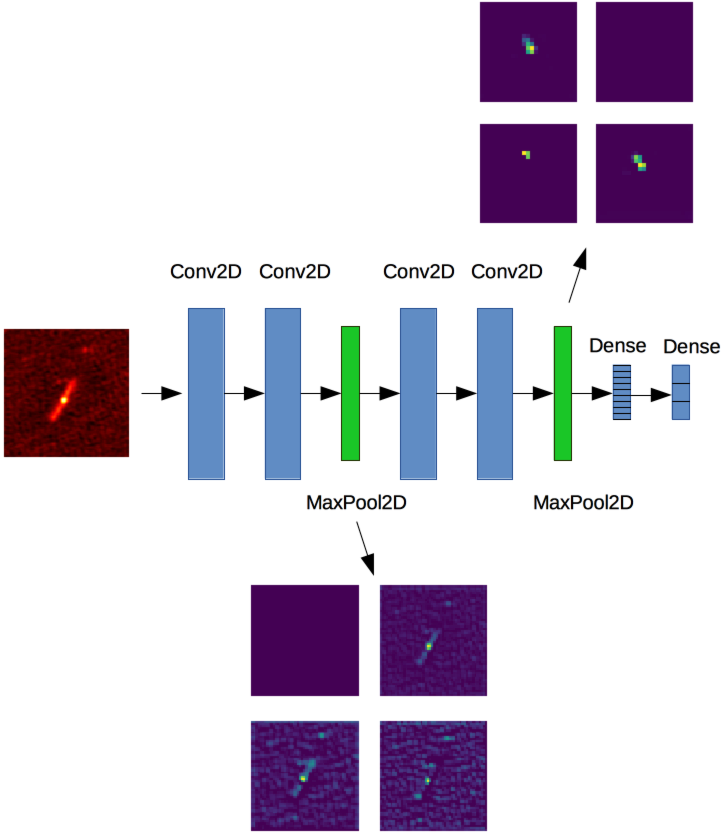}
    \caption{The ConvNet-4 architecture. The input to the network is a 100x100x3 image. Showing an example input image with features detected at the second and fourth convolutional layers, after pooling, at the end of training (50 epochs). We show 4 feature maps for each of the two outputs.}
    \label{fig:conv_feature_Avg. F1s}
\end{figure}

\begin{figure}
    \includegraphics[width=83mm,scale=0.9]{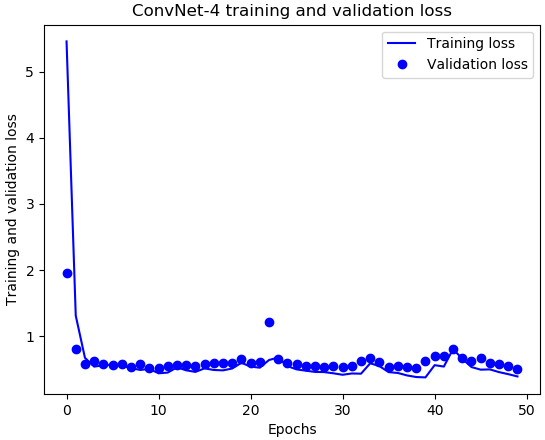}
    \caption{The training and validation losses for a single run with the ConvNet-4 architecture using the cross-entropy loss, with 2301 (79\%) samples for training and 600 (21\%) samples for testing.}
    \label{fig:convnet_total_loss}
\end{figure}

Examples of detected features in the ConvNet-4 model at the output of the second and fourth convolutional layers, after max pooling are shown in Figure~\ref{fig:conv_feature_Avg. F1s}. The training and validation losses for a single run with the ConvNet-4 architecture are shown in Figure \ref{fig:convnet_total_loss}. 

The use of a more complex architecture (ConvNet-8 compared to ConvNet-4) appears to improve the classification metrics (Avg. Recall = 91.2 compared to 88.7 respectively).

\subsubsection{CapsNet model}

Each epoch consisting of 2301 training samples takes approximately 3.4 minutes for the default model, 14 seconds for the inc. filtersize model and 3.5 minutes for the inc. decoder model. The faster time for the inc. filtersize model is due to the fact that the feature maps are moved across the image by 4 pixels (stride of 4) in the first convolutional layer as opposed to using a stride of 1, therefore the feature maps are able to scan through the image faster. 

\begin{table}
	\centering
	\caption{The average metrics (in percentages) across each of the classes in the original LOFAR dataset, for the default CapsNet model. Five runs were done in total, using 600 samples in the test set.}
	\label{tab:LOFAR_caps_1}
	\begin{tabular}{lcccr}
		\hline
		Class & Precision & Recall & F1 score & Accuracy \\
		\hline
		\textbf{(1)} & & & \\
		Unres. & 92.7 $\pm$ 1.4  & 95.7 $\pm$ 0.7  & 94.2 $\pm$ 1.0 & 93.4 $\pm$ 1.2 \\ 
		FRI & 78.3 $\pm$ 3.1 & 87.7 $\pm$ 1.6  & 82.7 $\pm$ 1.1 & 86.3 $\pm$ 1.3 \\
		FRII & 66.6 $\pm$ 5.1  & 35.0 $\pm$ 13.0  & 43.1 $\pm$ 12.7 & 88.2 $\pm$ 0.8 \\
		\hline
		\hline
		Avg. & 84.0 $\pm$ 1.3  & 84.7 $\pm$ 1.5 & 83.2 $\pm$ 2.6 & 90.1 $\pm$ 1.2 \\
		\hline
		\textbf{(2)} & & & \\
		Unres. & 96.4 $\pm$ 0.6 & 96.4 $\pm$ 0.9 & 96.4 $\pm$ 0.2 & 96.1 $\pm$ 0.2 \\
		FRI & 85.5 $\pm$ 1.4  & 90.2 $\pm$ 0.2  & 87.8 $\pm$ 0.7 & 90.7 $\pm$ 0.6 \\ 
		FRII & 75.8 $\pm$ 1.8  & 64.2 $\pm$ 0.5  & 69.6 $\pm$ 1.4 & 92.3 $\pm$ 0.4 \\ 
		\hline
		\hline
		Avg.& 89.7 $\pm$ 0.5  & 89.9 $\pm$ 0.5  & 89.7 $\pm$ 0.5 & 93.7 $\pm$ 0.3 \\ 
		\hline
		\textbf{(3)} & & & \\
		Unres. & 97.3 $\pm$ 0.5  & 98.1 $\pm$ 0.1 & 97.7 $\pm$ 0.3 & 97.5 $\pm$ 0.3  \\ 
		FRI & 90.9 $\pm$ 0.7 & 88.4 $\pm$ 0.8 & 89.6 $\pm$ 0.6 & 92.5 $\pm$ 0.5 \\ 
		FRII & 72.0 $\pm$ 2.6 & 75.2 $\pm$ 3.3 & 73.6 $\pm$ 2.8 & 92.7 $\pm$ 0.8 \\ 
		\hline
		\hline
		Avg.& 91.6 $\pm$ 0.7  & 91.5 $\pm$ 0.7  & 91.5 $\pm$ 0.7 & 95.0 $\pm$ 0.4 \\ 
        \hline
        \textbf{(4)} & & & \\
		Unres. & 98.4 $\pm$ 0.1 & 98.3 $\pm$ 0.1 & 98.3 $\pm$ 0.0 & 98.3 $\pm$ 0.0  \\ 
		FRI & 92.0 $\pm$ 0.6  & 91.3 $\pm$ 1.2 & 91.7 $\pm$ 0.5 & 93.9 $\pm$ 0.4 \\ 
		FRII & 80.4 $\pm$ 2.4  & 82.3 $\pm$ 1.8 & 81.2 $\pm$ 1.1 & 94.9 $\pm$ 0.4 \\ 
		\hline
		\hline
		Avg.& 93.7 $\pm$ 0.3  & 93.6 $\pm$ 0.4  & 93.6 $\pm$ 0.3 & 96.2 $\pm$ 0.2 \\ 
        \hline
	\end{tabular}
\end{table}

\begin{table}
	\centering
	\caption{The average metrics (in percentages) across each of the classes in the original LOFAR dataset, for the inc. filtersize CapsNet model. Five runs were done in total, using 600 samples in the test set.}
	\label{tab:LOFAR_caps_2}
	\begin{tabular}{lcccr}
		\hline
		Class & Precision & Recall & F1 score & Accuracy \\
		\hline
		\textbf{Orig.} & & & \\
		Unres. & 89.6 $\pm$ 0.7 & 94.2 $\pm$ 0.3  & 91.8 $\pm$ 0.5 & 90.8 $\pm$ 0.5 \\ 
		FRI & 80.4 $\pm$ 2.5 & 79.6 $\pm$ 2.9  & 79.9 $\pm$ 0.1 & 85.0 $\pm$ 0.5 \\
		FRII & 63.2 $\pm$ 6.4  & 50.5 $\pm$ 10.8  & 54.2 $\pm$ 6.7 & 88.4 $\pm$ 0.2 \\
		\hline
		\hline
		Avg. & 82.7 $\pm$ 0.5  & 83.0 $\pm$ 0.5 & 82.5 $\pm$ 1.1 & 88.4 $\pm$ 0.4 \\
		\hline
	\end{tabular}
\end{table}

\begin{table}
	\centering
	\caption{The average metrics (in percentages) across each of the classes in the original LOFAR dataset, for the inc. decoder CapsNet model. Five runs were done in total, using 600 samples in the test set.}
	\label{tab:LOFAR_caps_3}
	\begin{tabular}{lcccr}
		\hline
		Class & Precision & Recall & F1 score & Accuracy \\
		\hline
		\textbf{Orig.} & & & \\
		Unres. & 90.6 $\pm$ 2.7  & 95.0 $\pm$ 0.8 & 92.7 $\pm$ 1.8 & 91.6 $\pm$ 2.2  \\ 
		FRI & 75.1 $\pm$ 2.5 & 87.8 $\pm$ 1.9  & 80.9 $\pm$ 2.2 & 84.5 $\pm$ 1.9  \\
		FRII & 65.8 $\pm$ 2.9 & 22.7 $\pm$ 9.0  & 32.3 $\pm$ 10.7 & 87.4 $\pm$ 0.8 \\
		\hline
		\hline
		Avg. & 81.6 $\pm$ 2.0  & 82.7 $\pm$ 2.1 & 80.3 $\pm$ 3.0 & 88.5 $\pm$ 1.9  \\
		\hline
	\end{tabular}
\end{table}

\begin{figure}
    \includegraphics[width=83mm,scale=0.9]{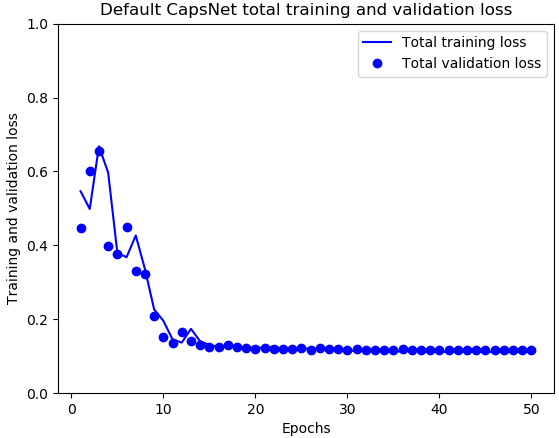}
    \caption{The training and validation losses for a single run with the default capsule network architecture, using the margin loss as defined in Equation~\ref{eq:5}, with 2301 (79\%) samples for training and 600 (21\%) samples for testing. The total loss is obtained by adding the capsule network loss to the decoder weight multiplied by the decoder loss.}
    \label{fig:capsnet_total_loss}
\end{figure}

Examples of detected features at the PrimaryCaps layer, prior to the reshape and squashing functions are shown in Figure~\ref{fig:capsnet_architecture} for the default model. Figure~\ref{fig:capsnet_total_loss} shows the training and validation loss curve for the default model. Table~\ref{tab:LOFAR_caps_1} shows that the default model attains higher overall metrics compared to the other two CapsNet models (although this is not always significant). 

The inc. filtersize model, that was designed with larger filters to capture more extended emission, for the most part performs as well as the default model and the metrics for the FRIIs are improved. However, they tend to be lower for the Unresolved and FRI classes, which make up the majority of samples. The results are shown in Table~\ref{tab:LOFAR_caps_2}. 

The inc. decoder model, which uses a more complex decoder, performs as well as the default model in the metrics for the Unresolved and FRI classes.  However, it performs worse overall for the FRIIs, as shown in Table~\ref{tab:LOFAR_caps_3}. This may be due to the more complex decoder confusing radio emission from the FRIIs with noise. 

As the default CapsNet model performs better overall compared to the other two CapsNet models, it is chosen as the basis of comparison against the two ConvNet models across the original {\sc fits} and masked 4rms sigma-clipped datasets. 

The default CapsNet model still performs significantly worse compared to the two ConvNets, as it is beyond both their 95\% confidence intervals, across all metrics. The variability in metrics is higher for the original dataset compared to that of the two ConvNets, as is evident in the generally increased confidence intervals of the CapsNet model, in Table \ref{tab:LOFAR_caps_1}, particularly for the FRIIs. 

\begin{figure}
    \includegraphics[width=83mm,scale=0.9]{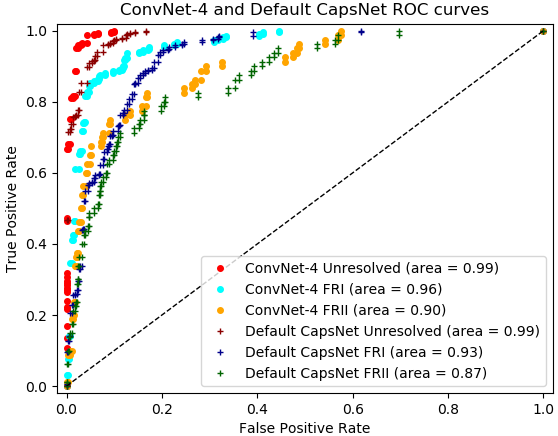}
    \caption{ROC curves for both a single run with the default CapsNet model and the ConvNet-4 model. The curves
    show that ConvNet-4 outperforms the default CapsNet across all the classes.}
    \label{fig:capsnet_conv_roc_curve}
\end{figure}

\begin{figure}
    \includegraphics[width=83mm,scale=0.9]{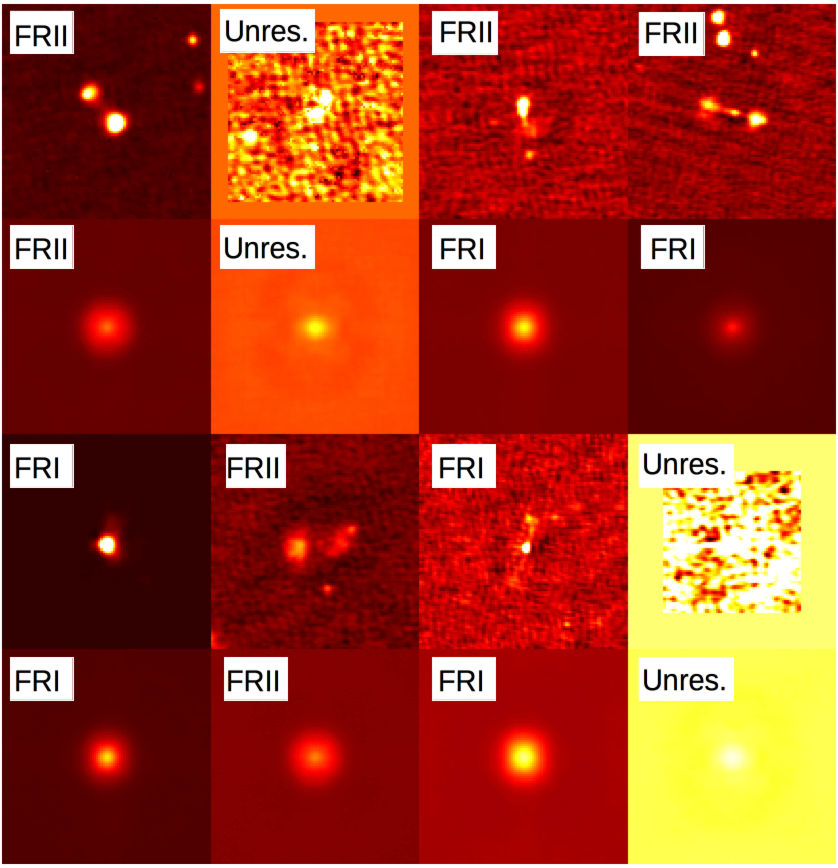}
    \caption{The real and reconstructed images using the default capsule network setup when training on the original and augmented images, annotated with the corresponding labels. The top row shows the real images, the second row shows the corresponding reconstructions. The third row shows the real images and the final row shows the reconstructions. The decoder always detects that there is an object in the centre of the image, however it is unable to reconstruct the object accurately. Based on the reconstruction, we see that CapsNet is determining class membership based on the characteristics of the sphere in the centre.} 
    \label{fig:real_and_recon}
\end{figure}

Figure~\ref{fig:capsnet_conv_roc_curve} shows the Receiver Operating Characteristic (ROC) curves across the default capsule network and ConvNet-4. ROC curves plot the true positive rate (recall) against the false positive rate.

In a first attempt to use the default CapsNet model (containing 58M free parameters), we observed a clear overfitting, owing to the large number of free parameters compared to the number of training images. Despite this, the model still achieved very similar results to the models using many fewer parameters quoted in the current work.

\begin{figure}
    \includegraphics[width=83mm,scale=0.9]{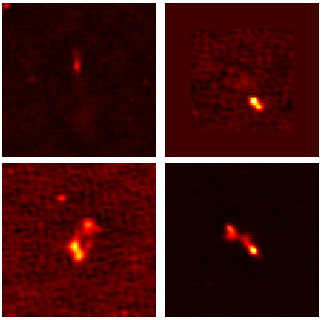}
    \caption{Examples of radio galaxies having probabilities greater than 0.5 in more than two classes in the default CapsNet architecture, that are also incorrectly predicted. The labels and predictions from left to right, top to bottom are [FRI,Unres.,FRII,FRII] and [FRII,FRII,FRI,FRI] respectively. These sources are labeled as (1,2,3,4) in Table \ref{tab:Uncertain_capsnet_preds}.}
    \label{fig:capsnet_misclassified}
\end{figure}

Figure \ref{fig:capsnet_misclassified} shows four examples of radio galaxies in which the probabilites are greater than 50\% across 2 classes, that the CapsNet could therefore not reliably classify. There are a total of 55 out of 600 (9.2\%) such cases. Table \ref{tab:Uncertain_capsnet_preds} shows the CapsNet probability vector across the four examples. In Source 1, CapsNet gives similar probabilities between the FRI and FRII classes, which could be because the source is quite faint, therefore it is having trouble extracting the morphology. Source 2 is predicted more confidently as an FRII compared to an Unresolved source, perhaps because it appears as though it has two lobes close together. Sources 3 and 4 are labeled as an FRII, however the CapsNet predicts them more confidently as an FRI compared to an FRII, as it may not detect the lobes.

\begin{table}
	\centering
	\caption{The labels and corresponding probability vector of the default CapsNet network predictions, using four examples of sources shown in Figure \ref{fig:capsnet_misclassified}, having probabilities greater than 50\% across two classes.}
	\label{tab:Uncertain_capsnet_preds}
	\begin{tabular}{lcr}
	    \hline
		Source & Label  & CapsNet prediction \\
		\hline
		1 & FRI & 41\% Unres., 50\% FRI, 51\% FRII \\
		2 & Unres. & 51\% Unres., 36\% FRI, 62\% FRII \\
		3 & FRII & 34\% Unres., 59\% FRI, 57\% FRII \\
		4 & FRII & 16\% Unres., 72\% FRI, 70\% FRII \\
		\hline
	\end{tabular}
\end{table}

\subsection{LOFAR original and augmented images}
\label{sec:orig+aug}
We augmented the images with translation, rotation and flipping as outlined in Section~\ref{sec:augmentation}, keeping the distribution of FRI and FRII sources the same as in the original dataset. Table~\ref{tab:Number_sources_initial_augmented} gives the number of original and augmented images. There are again 79\% and 21\% of the original samples used in training and testing respectively.

\subsubsection{ConvNet-4 and ConvNet-8}

We applied both ConvNet-4 and ConvNet-8 models to the original and augmented dataset, with the results shown in Table~\ref{tab:LOFAR_conv_1} and Table~\ref{tab:LOFAR_conv_2}. The overall metrics are significantly better (Avg. Recall = 93.4 and 94.3) than was observed when the same model was used on the original images (Avg. Recall = 88.7 and 91.2 for ConvNet-4 and ConvNet-8 respectively), therefore both models benefit from data augmentation. The confidence intervals are also usually reduced.

Although the classification metrics remain the poorest for the FRII class, they improved the most when using the augmented data, despite the fact that there were more examples of FRIs.

A confusion matrix is provided in Table \ref{tab:Confusion_matrix1} for the ConvNet-4 model, to see the numbers of samples that are both correctly and incorrectly predicted.

\begin{table}
	\centering
	\caption{Confusion matrix for a single run with the ConvNet-4 architecture, after training on the original and augmented images. The predictions are along the columns and the labels are along the rows.}
	\label{tab:Confusion_matrix1}
	\begin{tabular}{l | c | c | c | r}
	    \hline
		 & Unres. & FRI & FRII & Total \\
		\hline
		Unres. & 294 & 6 & 1 & 301 \\
		FRI & 3 & 202 & 14 & 219 \\
		FRII & 5 & 12 & 63 & 80 \\
		\hline
		Total & 302 & 220 & 78 & 600 \\
		\hline
	\end{tabular}
\end{table}

\subsubsection{CapsNet}

The best-performing capsule network (the default model) was used to see whether an improvement in overall metrics could be obtained when using augmented images in addition to the original images. The results are shown in Table ~\ref{tab:LOFAR_caps_1}. The confusion matrix for a single run with the default CapsNet architecture, after training on the original and augmented images, is given in Table \ref{tab:Confusion_matrix2}.

The classification metrics are significantly improved when using the augmented images (Avg. Recall = 89.3 with augmentation, compared to Avg. Recall $= 84.2 \pm 0.2$ without), therefore the capsule network also benefits from training on additional images. Despite the fact that capsule networks output a vector describing the properties of images across the classes and aim to extract the underlying patterns, they still benefit from the use of additional augmented images, for the {\sc fits} file dataset. The noise in the images could be preventing the network from seeing the underlying morphology in the signal, and there is an insufficient number of images available across the classes, hence improved results are observed when more examples are provided. Despite CapsNet benefiting from augmentation, the classification metrics are still significantly lower compared to when augmentation is applied to the two ConvNets.

Figure~\ref{fig:real_and_recon} shows the real and reconstructed images for a single run of the default CapsNet model when training on the original and augmented images. The labels match the predictions with the exception of the third and fourth images in the top two rows, where the true labels are FRIIs but the predictions are FRIs. The reconstructions of the images are innaccurate, giving the appearance that CapsNet is determining class membership based on the blurriness of the reconstructed spheres. The images in the \textquoteleft Unresolved' class are represented as concentrated spheres, FRIs are less concentrated, blurrier spheres, and FRIIs are the most diffuse. The inaccuracy of the reconstructions is most likely due to the fact that CapsNet appears to have trouble distinguishing signal from noise. Despite this, the average metrics are still above 89\% when training on the original and augmented images, as it does not appear to be necessary to have accurate reconstructions to determine class membership. Overall, the FRII source predictions appear to be the most affected by the noise level and/or potential unassociated emission in the images; since the reconstructions tend to be blurrier spheres with only one component, they become confused with FRIs and FRIIs, as FRIIs can have either both lobes being connected, as well as disconnected.

Similar to what was observed in the ConvNet architectures, the metrics across the FRII class are the poorest. However, after training with the original and augmented images, the FRII metrics improved the most. The FRII class has the fewest examples of images compared to the other two classes.

Despite the use of image augmentation, it is likely that the number of original training samples available is insufficient to train a capsule network.

\begin{table}
	\centering
	\caption{Confusion matrix for a single run with the default CapsNet architecture, after training on the original and augmented images. The predictions are along the columns and the labels are along the rows.}
	\label{tab:Confusion_matrix2}
	\begin{tabular}{l | c | c | c | r}
	    \hline
		 & Unres. & FRI & FRII & Total \\
		\hline
		Unres. & 289 & 12 & 0 & 301 \\
		FRI & 4 & 198 & 17 & 219 \\
		FRII & 6 & 24 & 50 & 80 \\
		\hline
		Total & 299 & 234 & 67 & 600 \\
		\hline
	\end{tabular}
\end{table}

\subsection{Sigma-clipped images}
\label{sec:Sigma-clipped}
In order to test whether the CapsNet performance could be improved by removing noise and the occasionally unassociated emission, we used the sigma clipped images that mask out pixels below 4rms. A flood-filling algorithm and masking techniques have additionally been applied to the dataset to identify and connect associated emission (Mingo et al. in prep). We analyse the results obtained from using the original sigma-clipped images, as well as both the original and augmented images.

The performance of both ConvNets is significantly improved as shown in Tables \ref{tab:LOFAR_conv_1} and \ref{tab:LOFAR_conv_2} (Avg. Recall = 91.9\% compared to 88.7\% for ConvNet-4, 94.9\% compared to 91.2\% for ConvNet-8) when using the original sigma clipped images, compared to using the original {\sc fits} files that includes noise and potential unassociated sources. The use of the original sigma clipped images is significantly worse compared to using the original and augmented {\sc fits} images for the ConvNet-4 model (Avg. Recall = 91.9\% compared to 93.4\%), and is not significantly better for the ConvNet-8 model. The inclusion of augmented images on the sigma-clipped dataset appears to benefit the ConvNet-4 model more compared to the ConvNet-8 model.

The performance of CapsNet is significantly improved as shown in Table \ref{tab:LOFAR_caps_1} when using the sigma-clipped original images (Avg. Recall = 91.5\% compared to 84.7\% with the original {\sc fits} images, and compared to 89.9\% with the original and augmented {\sc fits} images). However, CapsNet still performs worse compared to both ConvNet-4 and ConvNet-8. The use of image augmentation on the sigma-clipped images appears to improve the performance (Avg. Recall = 93.6\% compared to 91.5\%) The confidence intervals are also generally smaller compared to when the {\sc fits} images are used, therefore the performance is slightly more stable.

The use of the sigma clipped and masked arrays is also significantly better than using the {\sc fits} images, when comparing the performance within the original, and the original and augmented datasets, across both ConvNet models and CapsNet models. Therefore, none of the deep learning models can be trained to be completely robust to noise and potentially unassociated emission.

\begin{figure}
    \includegraphics[width=83mm,scale=0.9]{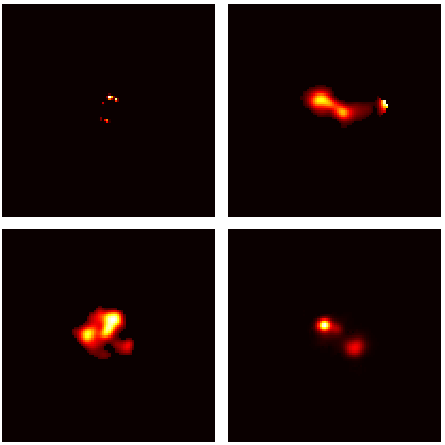}
    \caption{Examples of incorrectly classified radio galaxies from the 4rms sigma-clipped dataset using the ConvNet-8 layer architecture. The labels and predictions from left to right, top to bottom are [Unres.,FRII,FRII,FRI] and [FRI,FRI,FRI,FRII] respectively. The top left image appears to have too few pixels to be reliably classified, thus belonging to the \textquoteleft unresolved' class, however the remaining three may have been misclassified by the automated algorithm.}
    \label{fig:misclassified}
\end{figure}

In considering the results of one particular run with the ConvNet-8 model, out of 600 test samples, there are 20 where the predictions do not match the labels. Figure \ref{fig:misclassified} shows four such examples of images from the 4rms sigma-clipped dataset. Upon inspection of all the incorrectly predicted radio galaxies using the ConvNet-8 model, all 12 images that have been labelled as an FRII are predicted to be an FRI. Out of 3 images labeled as \textquoteleft Unresolved', two are predicted to be an FRI and one is predicted to be an FRII. The remaining 5 images labelled as FRI are predicted to be FRIIs. The wrongly classified galaxies mostly appear to have an ambiguous morphology and therefore it could be argued that they are mis-classified by the automated algorithm used to label them (see Section \ref{sec:Classifications} and Mingo et al. (in prep.)). For example, the top right and bottom left panels in Figure \ref{fig:misclassified} do not appear to be a representative examples of an FRII, and the bottom right panel appears more as an FRII, whereas it is labeled as an FRI. 

We note that the larger the proportion of sources that are mis-classified by the automated algorithm, the more difficult it will be for the models to learn.

\subsection{Additional results}

This section summarises other convolutional and capsule network architectures as well as parameters, that were tried. These include transfer learning, the application of early stopping and comparison of results with similar work.

\subsubsection{ConvNet models}

We also wanted to test the performance of a simple purely convolutional architecture using 5 layers (with no intermediate dense layer following the convolutions). The purpose of these dense layers is to help model complex global patterns in the data. The metrics were significantly lower compared to those of both ConvNet models, as shown in Table~\ref{tab:five_conv_layer}. Therefore, at least one intermediate dense layer could be necessary for optimal performance in convolutional networks. We also tested an architecture using 4 convolutional with no pooling layers, and found the results to be inferior compared to using the ConvNet-4 model. Therefore, the use of pooling is appears to be advantageous in the current dataset, perhaps because it allows more degrees of freedom for the morphology within classes.

\begin{table}
	\centering
	\caption{The average metrics (in percentages) across each of the classes in the (2) original and augmented LOFAR dataset using a 5 convolutional layer model with no intermediate dense layers. Five runs were done in total, using 600 samples in the test set.}
	\label{tab:five_conv_layer}
	\begin{tabular}{lcccr}
		\hline
		Class & Precision & Recall & F1 score & Accuracy \\
		\hline
		\textbf{(2)} & & & \\
		Unres. & 97.7 $\pm$ 0.6  & 96.9 $\pm$ 0.9  & 97.3 $\pm$ 0.3 & 97.2 $\pm$ 0.3  \\ 
		FRI & 88.4 $\pm$ 1.7 & 92.8 $\pm$ 1.9  & 90.5 $\pm$ 0.1 & 92.8 $\pm$ 0.1   \\
		FRII & 77.8 $\pm$ 2.9 & 69.0 $\pm$ 2.5  & 73.0 $\pm$ 1.3 & 93.1 $\pm$ 0.4  \\
		\hline
		\hline
		Avg. & 91.6 $\pm$ 0.2 & 91.7 $\pm$ 0.1 & 91.6 $\pm$ 0.1 & 95.0 $\pm$ 0.1  \\
		\hline
	\end{tabular}
\end{table}

\subsubsection{CapsNet models}

Other variations on capsule network models included stacking two convolutional layers instead of one, using 90\% training data and 10\% testing data, using an ensemble of capsule network models, increasing the number of routing iterations, decreasing the filter size, changing the batch size, adjusting the learning rate, using different activation functions, applying dropout, pooling and using a combination of increased filter sizes together with a more complex decoder, all which resulted in similar or worsened performance metrics. The only possible improvement could be the use of a larger sample of original training images.

\subsubsection{Transfer learning}

Transfer learning \citep{Pratt_etal1991} involves applying the knowledge from one trained neural network to help another learn a related task. In the deep learning context, weights are typically pre-loaded from a network trained on a large dataset with many classes to another unseen dataset. 

We used the Inception ResNet model v2 \citep{Szegedy_etal2016}, which combines Inception and Residual network architectures. An inception network consists of a convolutional network using filters of various sizes and pooling within the same layer, and a residual network utilises skip connections between convolutional layers if the classification accuracy becomes saturated with the subsequent stacking of layers. The Inception ResNet model is trained on the ImageNet dataset \citep{Deng_etal2009}, to classify over 14M images into 1000 categories. Although the nature of the ImageNet dataset is different to the radio galaxy images, pre-loading weights from a network trained with such a dataset is better than initialising the weights from a random distribution.

To use the pre-trained ResNet model in Keras requires images of size of at least 139x139 pixels. As such we padded our images with zeros for 20 pixels along the horizontal and vertical directions, resulting in images of 140x140 pixels.

The pre-trained ResNet model is applied to the LOFAR original and augmented {\sc fits} images, to verify whether the classification metrics could be improved from those of our other models. The results in Table~\ref{tab:LOFAR_transfer_learn} show that the classification metrics are not significantly better (Avg. Recall = 94.5\%) compared to when training on the same set of images from randomly initialised weights with the ConvNet-8 architecture (Avg. Recall = 94.3\%). The metrics are significantly better than for the ConvNet-4 architecture (Avg. Recall = 93.4\%). Optimal results are still obtained when using the sigma clipped dataset, where noise and potentially unassociated sources are removed. 

We note that the results obtained with transfer learning may be improved if there is a neural network trained on a similar astronomical classification task from which pre-trained weights can be loaded. A successful implementation of transfer learning in classifying optical galaxy morphology is in \citet{Dominguez_etal2019}, and most recently in radio galaxy morphology classification \citep{Tang_etal2019}.

The pre-trained network converges faster; ConvNet-4 required 40 epochs of training to reach the optimal validation accuracy as opposed to 30 epochs for the transfer learning model, when averaged over five runs.

\begin{table}
	\centering
	\caption{The average metrics (in percentages) across each of the classes in the (2) original and augmented LOFAR dataset, for the transfer learning model. Five runs were done in total, using 600 samples in the test set.}
	\label{tab:LOFAR_transfer_learn}
	\begin{tabular}{lcccr}
		\hline
		Class & Precision & Recall & F1 score & Accuracy \\
		\hline
		\textbf{(2)} & & & & \\
		Unres. & 98.7 $\pm$ 0.2 & 98.3 $\pm$ 0.4  & 98.5 $\pm$ 0.2 & 98.4 $\pm$ 0.2 \\ 
		FRI & 91.8 $\pm$ 0.5 & 95.0 $\pm$ 0.4 & 93.4 $\pm$ 0.2 & 95.0 $\pm$ 0.2 \\
		FRII & 85.4 $\pm$ 1.2 & 78.7 $\pm$ 2.7 & 81.9 $\pm$ 1.2 & 95.3 $\pm$ 0.2\\
		\hline
		\hline
		Avg. & 94.4 $\pm$ 0.2 & 94.5 $\pm$ 0.2 & 94.4 $\pm$ 0.2 & 96.8 $\pm$ 0.1 \\
		\hline
	\end{tabular}
\end{table}

\subsubsection{Early stopping}
\label{sec:Early_stopping}
We also experimented with applying early stopping in the training of both the Capsnet and ConvNet models. The implementation was such that if the validation accuracy did not improve for 10 subsequent epochs, training was stopped and the metrics on the test set were calculated. However, we found the performance to be the same for the ConvNet model, and worse for the CapsNet model, compared to when training for a pre-defined number of 50 epochs (results not shown). In a work focused on the usage of early stopping, \citet{Prechelt_2012} used a mix of more than 1000 training runs across 12 different problems and 24 different architectures and concluded that slower stopping criteria allow for $\approx$ 4\% average improvement in generalisation, at a cost of around a factor of four longer in training time. 

\subsubsection{Recent similar work}

Recently, \citep{Katebi_etal2018} applied a capsule network to classify optical galaxies based on morphology, using the classes of spiral, elliptical and star/artefact. They find that their capsule network classification accuracy surpasses that of their baseline convolutional network (98.77\% versus 96.96\% respectively). The capsule network architecture has over 124M parameters, for a total of 61,578 images. In contrast, our best-performing capsule network uses just over 4M parameters with up to 15,936 images using the original and augmented dataset. 

We note that the difference in morphology between their classes is starker than in our case. Additionally, the optical images show a much better contrast between object and background, where noise is less prominent. The optical galaxy classifications were crowd-sourced, whereas our labels originated from an automated algorithm which comes with some limitations, as outlined in Section \ref{sec:Classifications}. The radio emission also produces sparser images compared to the optical galaxy images.

It is difficult to compare their work to ours as the number of images in each of their 3 classes is unknown. Hence, it is uncertain whether the classification accuracy is the best discriminator to use between the models \citep{ Hossin_Sulaiman_2015}. Other classification metrics are not provided, such as precision and recall, which may be more powerful in discriminating models. There is also no indication of variability between runs, as well as the degree of overfitting in the networks during training.

\section{Conclusions}
\label{sec:conclusions}

This paper explored two deep learning approaches in the classification of radio data from the LoTSS HETDEX field across three classes of radio galaxies: Unresolved sources, FRI and FRII galaxies. The labels were generated using an automated algorithm, which used a catalogue of sources from the LoTSS DR1 source catalogue with optical IDs and associations \citep{Williams_etal2019}. The radio galaxies belonging to the FRI and FRII classes were additionally cross-checked to eliminate galaxies in which the radio emission is likely to be dominated by star formation \citep{Hardcastle_etal2019}. Despite the classifications being generated using masked images that remove potentially unassociated sources and emission below 4rms from the images, one of our aims was to test how robust our deep learning algorithms could be when such effects were present.

We tested the performance of a four and eight layer convolutional neural network (ConvNet-4 and ConvNet-8) against various architectures of capsule networks (CapsNet), using the precision, recall, F1 score and accuracy, to evaluate the performance of the models. Python code implementing v1.0 of the algorithms can be obtained from github\footnote{https://github.com/vlukic973/RadioGalaxy\_Conv\_Caps}. Automated classifications of LoTSS sources obtained with the algorithms will be presented in a future paper (Mingo et al., in preparation). 

The first CapsNet model explored was the default model, a simplified architecture of the original model designed for the MNIST dataset, the second used larger filter sizes in the first convolutional layer and Primary capsule layer, and a larger stride in the convolutional layer. The third model used a more complex decoder and a higher loss for the decoder weight. The second and third models were designed to better account for the increased complexity of the data. Four different sets of data were used to train and test the two ConvNets and the variations on CapsNet architectures: (i) using the original {\sc fits} images only, (ii) original and augmented {\sc fits} images, (iii) the original masked arrays that remove emission below 4rms and potential unassociated sources and (iv) original and augmented masked 4rms arrays.

We found that the optimal CapsNet performance was obtained when using the default model, in terms of the overall classification metrics.

The results showed that the ConvNet architectures  always exceeded the performance of the chosen CapsNet model, and ConvNet-8 always performed better compared to ConvNet-4, most likely because the ConvNet-8 model has twice the number of convolutional layers and parameters as ConvNet-4, therefore it is able to extract higher-dimensional features that are particular to each class.

The use of transfer learning on the original and augmented images achieved the same results as ConvNet-8. The performance of all deep learning models was optimised when using the 4rms sigma clipped numpy array, which is expected as the noise and potential unassociated emission is removed. Some observations of differences in results between using ConvNet and CapsNet architectures and the likely reasons are as follows:

\begin{itemize}
    \item As CapsNet tends to capture and preserve the relative location of features in the images, it is not as successful in distinguishing signal from noise, or dealing with the presence of potentially unassociated emission, as the ConvNet architectures 
    \item The use of pooling in the ConvNet architectures generally appears to be advantageous in two respects: (i) increased likelihood that noise and potential unassociated sources will be filtered out, (ii) allowing more degrees of freedom for variability in morphology within the classes, when the undesirable effects have been removed through use of the 4rms dataset
    \item The removal of noise and potentially unassociated emission through the use of sigma-clipped and masked arrays improves the performance of both deep learning approaches, when considering the metrics within the original, and original and augmented datasets 
    \item The use of image augmentation appears to benefit both ConvNets and CapsNet, when using the {\sc fits} files, which contain the original radio emission.
\end{itemize}

The LoTSS survey is the first wide-area survey to contain such faint sources. It is sensitive to a larger range of source evolutionary states, and can also see structure on a wider range of spatial scales due to the combination of well-sampled UV coverage and long baselines. These features result in images having richer, more varied and sometimes ambiguous morphologies that are more difficult to categorise into distinct classes.

Across both deep learning algorithms, the \textquoteleft Unresolved' class is recovered most successfully, followed by the FRI class. The FRIIs tend to be the least well recovered. Although FRIs display morphological diversity as they can be straight or bent, FRIIs have two peaks of varying distances that may or may not be connected by extended emission with the host galaxy. Therefore, FRIs are more likely to contain a single connected component whereas FRII can contain either a single or two connected components. There are also fewer  examples of FRIIs in the dataset compared to FRIs. When we inspected some incorrectly predicted galaxies using the sigma-clipped dataset, we found the morphologies to be ambiguous in most cases, as shown in Figure \ref{fig:misclassified}.

Traditional convolutional neural networks generally contain pooling layers in their architecture in order to reduce the number of parameters. However, this can cause the relative locations of features within the image to degrade, which capsule networks are designed to preserve. Our results indicate that for the radio galaxy data in the current work, the performance of capsule networks is inferior to that of convolutional neural networks. This could be due to the number of original samples being insufficient to train the capsule network. Another reason may be that since they attempt to preserve the relative location of features, capsule networks appear to interpret noise as signal and introduce extra distortion into the image, as shown in Figure~\ref{fig:capsnet_architecture}. This aspect has proven to be most detrimental in the recovery of FRII sources, as they are more susceptible to the mingling of signal with noise due to the fact that they are comprised of either one or two components. Additionally, the FRII class contains the fewest examples of images.

In comparison with previous works that use convolutional neural networks to classify radio galaxy morphologies (\citet{Aniyan_Thorat2017}, \citet{Lukic_2018}, \citet{Wu_etal2018} and \citet{Alhassan_etal2018}), the current work explored the use of capsule networks, which are designed to preserve the hierarchical feature information in an image, and finds their performance to be inferior to that of standard convolutional network architectures. The data from the LOFAR LoTSS survey reveals fainter and more detailed emission compared to the data from the surveys which the previous works analysed, providing additional challenges for classification. As such, our findings hold for surveys having a comparable setup, provided they produce images with similar morphologies and noise profiles.

Based on the current results obtained, it appears that convolutional neural networks still hold as the deep learning technique that should be used for future surveys. They are also faster to train as they use fewer parameters. Capsule networks, in their present form, are generally slower and require further development to be made more robust to noisy real data, however the current performance may be improved by explicitly training them on cleaned data with various examples of morphologies present within each class.

There are several limitations that would need to be overcome to apply these methods to large samples, such as the need for ancillary data to separate star-forming galaxies. The exclusion cannot be performed based purely on the radio galaxy morphology. The classes should also be extended to encompass the hybrid sources, as well as other rare sources such as bent-tailed and double-double sources.

\section*{Acknowledgements}

We thank Huub Rottgering and the anonymous referee for useful comments. VL acknowledges support by the Deutsche Forschungsgemeinschaft (DFG) through grant SFB 676. PNB is grateful for support from STFC via grant ST/R000972/1. This paper is based on data obtained with the International LOFAR Telescope (ILT) under project codes LC2\_038 and LC3\_008. LOFAR \citep{van_Haarlem_etal2013} is the Low Frequency Array designed and constructed by ASTRON. It has observing, data processing, and data storage facilities in several countries, which are owned by various parties (each with their own funding sources), and which are collectively operated by the ILT foundation under a joint scientific policy. The ILT resources have benefited from the following recent major funding sources: CNRS-INSU, Observatoire de Paris and Universite d'Orleans, France; BMBF, MIWF-NRW, MPG, Germany; Department of Business, Enterprise and Innovation (DBEI), Ireland; NWO, The Netherlands; The Science and Technology Facilities Council (STFC), UK.












\bsp	
\label{lastpage}
\end{document}